\documentclass[prd,showpacs,aps,prd,nofootinbib,floatfix,amsmath,amssymb]{revtex4}
\usepackage{graphicx}
\usepackage{slashed}
\usepackage{textcomp}
\begin{document}

\makeatletter
\newbox\slashbox \setbox\slashbox=\hbox{$/$}
\newbox\Slashbox \setbox\Slashbox=\hbox{\large$/$}
\def\pFMslash#1{\setbox\@tempboxa=\hbox{$#1$}
  \@tempdima=0.5\wd\slashbox \advance\@tempdima 0.5\wd\@tempboxa
  \copy\slashbox \kern-\@tempdima \box\@tempboxa}
\def\pFMSlash#1{\setbox\@tempboxa=\hbox{$#1$}
  \@tempdima=0.5\wd\Slashbox \advance\@tempdima 0.5\wd\@tempboxa
  \copy\Slashbox \kern-\@tempdima \box\@tempboxa}
\def\FMslash{\protect\pFMslash}
\def\FMSlash{\protect\pFMSlash}
\def\miss#1{\ifmmode{/\mkern-11mu #1}\else{${/\mkern-11mu #1}$}\fi}
\makeatother

\title{One--loop nonbirefringent effects on the electromagnetic vertex in the Standard Model Extension}

\author{A. Moyotl$^{(a)}$, H. Novales--S\'anchez$^{( b )}$, J. J. Toscano$^{(a)}$, and E. S. Tututi$^{( c )}$}

\address{$^{(a)}$ Facultad de Ciencias F\'isico Matem\'aticas, Benem\'erita Universidad Aut\'onoma de Puebla, Apartado Postal 1152, Puebla, Puebla, M\'exico.\\ $^b$Divisi\'on de Ciencias e Ingenier\'ias, Universidad de Guanajuato Campus Le\'on, Loma del Bosque 103, Colonia Lomas del Campestre, 37150, Le\'on, Guanajuato, M\'exico.\\ $^{( c)}$ Facultad de Ciencias F\'isico Matem\'aticas, Universidad Michoacana de San Nicol\'as de Hidalgo, Avenida Francisco J. M\'ujica S/N, 58060, Morelia Michoac\'an, M\'exico.}

\begin{abstract}
Lorentz violation emerged from a fundamental description of nature may impact, at low energies, the Maxwell sector, so that contributions from such new physics to the electromagnetic vertex would be induced. Particularly, nonbirefringent $CPT$--even effects from the electromagnetic sector modified by the Lorentz-- and $CPT$--violating Standard Model Extension alter the structure of the free photon propagator. We calculate Lorentz--violating contributions to the electromagnetic vertex, at the one--loop level, by using a modified photon propagator carrying this sort of effects. We take the photon off shell, and find an expression that involves both isotropic and anisotropic effects of nonbirefringent violation of Lorentz invariance. Our analysis of the one--loop vertex function includes gauge invariance, transformation properties under $C$, $P$, and $T$, and tree--level contributions from Lorentz--violating nonrenormalizable interactions. These elements add to previous studies of the one--loop contributions to the electromagnetic vertex in the context of Lorentz violation in the photon sector. Finally, we restrict our analysis to the isotropic case and derive a finite contribution from isotropic Lorentz violation to the anomalous magnetic moment of fermions that coincides with the result already reported in the literature. 
\end{abstract}



\maketitle

\section{Introduction}
\label{intro}
Our current best fundamental description of nature, the Standard Model (SM), has overcome several accurate experimental tests, but is nowadays, however, considered to be an effective theory that does not include  all physical phenomena so far observed. It is then clear that there exist incentives to pursue more fundamental descriptions, and one path to achieve such a goal is by extending this low--energy description through slight deviations that add new elements providing an explanation of certain high--energy effects that could reveal hints on the most fundamental theory. Since symmetries are one of the main ingredients to build models, extensions of the SM gauge group and violations of symmetries associated to the discrete transformations $C$, $P$, and $T$ that keep invariance under $CPT$ have become the basement of several SM extensions. An interesting and relatively recent option was triggered by the breaking of Lorentz symmetry spontaneously generated~\cite{stringlv} in certain string--theory models and naturally arising~\cite{nclv} in noncommutative theories. The systematic low--energy treatment of such violations was crystallized in the Lorentz-- and $CPT$--violating Standard Model Extension (SME)~\cite{sme}, which is a general framework formulated within the effective field theory approach.

The SM and the SME share all dynamic variables and the gauge symmetry group as well. On the other hand, each of the SME terms involves a tensor that possesses Lorentz indices and whose components parametrize, at low energies, violation of Lorentz invariance that originates in a high--energy description. Lorentz violation can be related to either $CPT$--even or $CPT$--odd effects. It is very likely~\cite{K1} that the Lorentz--violating tensors of the SME are related to nonzero expectation values acquired by Lorentz tensors in a more fundamental theory after spontaneous breaking of Lorentz symmetry. The generality provided by the effective field theory approach allows one to study these Lorentz--violating new--physics effects independently of specific models, which has lead to several stringent bounds~\cite{KR} on these coefficients.

The most general version of the SME contains renormalizable couplings and nonrenormalizable ones~\cite{nrsme1,nrsme2,nrsme3,MNTT} as well, for no restriction from effective field theory to this respect exists. In the present paper, we consider the minimal SME, which is a subset of such general framework that includes only the Lorentz--violating terms that are renormalizable. We further restrict ourselves to the quantum electrodynamics sector with $CPT$--even Lorentz violation introduced in the photon sector, which corresponds to the Lagrangian~\cite{sme} ${\cal L}_{\rm QED}={\cal L}_\gamma+{\cal L}_{f\gamma}$, where
\begin{eqnarray}
{\cal L}_\gamma&=&-\frac{1}{4}F^{\mu\nu}F_{\mu\nu}-\frac{1}{4}(K_F)^{\mu\nu\alpha\beta}F_{\mu\nu}F_{\alpha\beta}-\frac{1}{2\xi}\,\partial^\mu A_\mu\,\partial^\nu A_\nu,
\label{cpteL}
\\ \nonumber \\
{\cal L}_{f\gamma}&=&\bar{\psi}(i\gamma^\mu D_\mu-m_f)\psi.
\end{eqnarray}
Here, $F^{\mu\nu}=\partial^\mu A^\nu-\partial^\nu A^\mu$  is the field strength tensor associated to the U(1) gauge field $A_\mu$, $\xi$ is the gauge fixing parameter, $D_\mu=\partial_\mu+ieA_\mu$ is the covariant derivative, and $\psi$ is a Dirac field with mass $m_f$ and the same charge as an electron. $CPT$--even violation of Lorentz invariance is introduced in this Lagrangian by the rank--4 tensor $(K_F)^{\mu\nu\alpha\beta}$, whose symmetries coincide with those of the Riemann tensor and which fulfills the double tracelessness condition $(K_F)_{\mu\nu}\hspace{0.001cm}^{\mu\nu}=0$. This leaves only nineteen independent components that can be classified~\cite{KM2} according to whether or not they play a role in birefringence. 

Interesting works involving radiative corrections produced by the SME exist~\cite{smerc1,smerc2,smerc3,smerc4,smerc5,smerc6,smerc7,smerc8,smerc9}. An important result that has been established~\cite{smerc5} is renormalizability of the one--loop contributions from the most general quantum electrodynamics sector of the minimal SME. Some calculations of radiative corrections from the SME to the $ff\gamma$ vertex, with $f$ representing a fermion, have been performed as well, including~\cite{MNTT,smerc3} the one--loop contributions that originate in the axial coupling of a Lorentz--violating background field to fermions. Our aim in the present paper is the calculation of nonbirefringent $CPT$--even Lorentz--violating contributions from the pure--photon sector of the minimal SME to the one--loop $ff\gamma$ vertex. This calculation was carried out before, in Ref.~\cite{smerc7}, by perturbatively inserting a Lorentz--violating two--point vertex function in the internal photon line of the loop diagram, in which all external particles were taken on shell. The main objective of such work was the derivation of a bound on isotropic Lorentz violation. In the present paper, we start from a modified photon propagator, which was derived in terms of two Lorentz--violating four--vectors in Ref.~\cite{CFGdS} and which involves nonbirefringent $CPT$--even and Lorentz--violating effects. We utilize the corresponding expression to calculate the one--loop contributions to the $ff\gamma$ vertex, which we derive at the lowest order in the corresponding Lorentz--violating parameter. Contrastingly to the investigation carried out in Ref.~\cite{smerc7}, our calculation is performed under the assumption that the external fermions are on shell, but the external photon is off shell. We provide the whole set of terms characterizing the one--loop $ff\gamma$ interaction generated by these modifications of the photon propagator, which includes both anisotropic and isotropic nonbirefringent effects. We find that, even though the contraction of the momentum of the external photon with the $ff\gamma$ vertex does not yield a simple Ward identity when the photon is taken on shell, a Ward--Takahashi identity is fulfilled. Though this identity indicates that our result is gauge invariant, we perform another proof that consists in constructing a set of nonrenormalizable terms governed by the electromagnetic gauge group and involving the nonbirefringent components of the $(K_F)^{\mu\nu\alpha\beta}$ tensor. The structure of the gauge invariant tree--level contributions from these effective terms to the $ff\gamma$ vertex matches that of our one--loop result, which consequently is gauge invariant.

We also take advantage of the nonrenormalizable terms to analyze the properties of the resulting one--loop $ff\gamma$ vertex under the discrete transformations $C$, $P$, $T$, $CP$, and $CPT$. At the first order in violation of Lorentz invariance, the parameters quantifying Lorentz violation in some terms of the resulting one--loop vertex form a Lorentz scalar. We note that terms carrying such Lorentz scalar are the only ones whose properties under space--time transformations and charge conjugation allow the generation of contributions to low--energy observables. For such reason, the contributions to low--energy quantities that arise exclusively involve this Lorentz scalar. In particular, we find a contribution from isotropic Lorentz violation to the anomalous magnetic moment of fermions that consistently coincides with the results of Ref.~\cite{smerc7} and update the bound obtained in that paper according to the latest improvement of the difference between the theoretical prediction of the SM contributions and the experimental data. 



The paper is organized as follows. In Sec.~\ref{QED} we define our framework while discussing some relevant issues about the photon sector of the minimal SME that we employ to perform this calculation. In Sec.~\ref{radiative}, we calculate the one--loop contributions to the $ff\gamma$ vertex function, showing the whole set of resulting couplings and discussing gauge invariance. We use a contribution to the magnetic dipole form factor to update the bound on the isotropic Lorentz violation parameter. Finally, our conclusions are provided in Sec.~\ref{conc}.

\section{The nonbirefringent Lorentz--violating photon propagator}
\label{QED}
As we commented in the Introduction, the manner in which we incorporate Lorentz--violating new--physics effects into the one--loop $ff\gamma$ vertex is through the photon propagator. Based on a couple of {\it ans\"atze}, the authors of Ref.~\cite{CFGdS} reported an expression of the photon propagator comprising effects of the nonbirefringent components of the tensor $(K_F)^{\mu\nu\alpha\beta}$, which we utilize to perform the loop calculation. In this section, we sketch the procedure posed in Ref.~\cite{CFGdS} to derive the corresponding exact tensorial expression, which we obtain in the general linear gauge, at the same time that we delineate our framework.


The $CPT$--even pure--photon sector of the minimal SME, Eq.~(\ref{cpteL}), can be written as
\begin{equation}
{\cal L}_\gamma=\frac{1}{2}A^\mu D_{\mu\nu}A^\nu,
\end{equation}
 where
\begin{equation}
\label{Dmn}
D_{\mu\nu}=g_{\mu\nu}\partial^2-2(K_F)_{\mu\alpha\beta\nu}\,\partial^\alpha\partial^\beta+\left( \frac{1}{\xi}-1 \right)\partial_\mu\partial_\nu.
\end{equation}
The first bound for the $CPT$--even and Lorentz violating modified Maxwell sector was given in Ref.~\cite{KM3}. Shortly after, in Ref.~\cite{KM2}, the same authors classified the components of the rank--4 tensor $(K_F)^{\mu\nu\alpha\beta}$ into birefringent and nonbirefrintent.
Since those works, this $CPT$--even, but Lorentz--violating, parametrization has been the subject of diverse works in both the theoretical~\cite{CFGdS,twlv1,twlv2,twlv3,twlv4,twlv5,twlv6,twlv7,twlv8,twlv9,twlv10,twlv11} and experimental~\cite{ewlv1,ewlv2,ewlv3,ewlv4} sides. There is a clear difference concerning the experimental sensitivity reached for each set of parameters, for the constraints established for the birefringent coefficients are more stringent than the current bounds restricting the nonbirefringent components by several orders of magnitude. Henceforth we concentrate solely in the nonbirefringent components, which can be parametrized, in terms of the rank--2 traceless symmetric tensor $k^{\mu\nu}=(K_F)^{\mu\alpha\nu}\hspace{0.001cm}_\alpha$, as~\cite{twlv4}:
\begin{equation}
(K_F)^{\mu\nu\alpha\beta}=\frac{1}{2}\left[ g^{\mu\alpha}k^{\nu\beta}-g^{\nu\alpha}k^{\mu\beta}+g^{\nu\beta}k^{\mu\alpha}-g^{\mu\beta}k^{\nu\alpha} \right].
\label{4tensor}
\end{equation}
The authors of Ref.~\cite{CFGdS} reduced the structure of the $k^{\mu\nu}$ tensor by proposing an expression of it in terms of two arbitrary four--vectors, $U^\mu$ and $V^{\mu}$. Such decomposition, whose explicit form is
\begin{equation}
k^{\mu\nu}=\frac{1}{2}\left( U^\mu V^\nu+U^\nu V^\mu \right)-\frac{1}{4}g^{\mu\nu}\,U\cdot V,
\label{2tensor}
\end{equation}
possesses the symmetry and null--trace properties of the original rank--2 tensor. Using Eqs.~(\ref{4tensor}) and (\ref{2tensor}), and denoting the momentum of the photon by $p$, the Fourier transform of Eq.~(\ref{Dmn}), which we represent by $\tilde{D}_{\mu\nu}$, reads
\begin{eqnarray}
\tilde{D}_{\mu\nu}&=& -\left[ p^2\left(1-\frac{1}{2}U\cdot V\right) +(p\cdot U)(p\cdot V)\right] g_{\mu\nu}-\left[ \frac{1}{\xi}-1-\frac{1}{2}U\cdot V \right] p_{\mu}p_{\nu}
\\ \nonumber&&
+\small{\frac{1}{2}}(p\cdot U)(p_{\mu}V_{\nu}+p_{\nu}V_{\mu})+ \frac{1}{2}(p\cdot V)(p_{\mu}U_{\nu}+p_{\nu}U_{\mu})-\frac{1}{2}p^2(U_{\mu}V_{\nu}+U_{\nu}V_{\mu}),
\end{eqnarray}
so that the propagator in momenta space, $\tilde{\Delta}^{\mu\nu}$, must fulfill $\tilde{\Delta}^{\mu\nu}\tilde{D}_{\nu\rho}=\delta^\mu\hspace{0.0001cm}_\rho$. Following Lorentz covariance and dimensional analysis,
a general expression of the photon propagator is constructed, which we show below:
\begin{eqnarray}
\tilde{\Delta}^{\mu\nu}&=&f_1\left[ g^{\mu\nu}-\frac{p^\mu p^\nu}{p^2}\right]+f_2\, \frac{p^\mu p^\nu}{p^2}+f_3\, U^{\mu}V^{\nu}+f_4\,U^{\nu}V^{\mu}+f_5\,p^{\mu}U^{\nu}+f_6\,p^{\nu}U^{\mu}
\nonumber \\ &&
+f_7\,p^{\mu}V^{\nu}+f_8\,p^{\nu}V^{\mu}+f_9\,U^{\mu}U^{\nu}+f_{10}\,V^{\mu}V^{\nu}.
\end{eqnarray}
The coefficients $f_i$, introduced in this manner, are unknown functions of the photon four--momentum and the four--vectors $U^\mu$ and $V^\mu$. The determination of such coefficients can be accomplished by performing all Lorentz--indices contractions in the propagator condition $\tilde{\Delta}^{\mu\nu}\tilde{D}_{\nu\rho}=\delta^\mu\hspace{0.0001cm}_\rho$, which yields a system of equations. The solution of such system provides the precise expressions of the $f_i$ coefficients and allows one to write the exact photon propagator as
\begin{equation}
\label{propagator}
\tilde{\Delta}^{\mu\nu}(  p  )=\frac{-i}{p^2(1+\Theta_{U,V})}\left[ g^{\mu\nu}+(\xi-1)\frac{p^\mu p^\nu}{p^2} \right]-i\,\Xi^{\mu\nu}( p ),
\end{equation}
where we have defined
\begin{eqnarray}
\Xi^{\mu\nu}( p )&=&\Bigg[\frac{\displaystyle F_1\,p^\mu p^\nu}{p^2}+F_2\left( U^\mu V^\nu+U^\nu V^\mu \right)+F_3\left( p^\mu U^\nu+p^\nu U^\mu \right)
\nonumber \\ &&
+F_4\left( p^\mu V^\nu+p^\nu V^\mu \right)+F_5\ U^\mu U^\nu +F_6V^\mu V^\nu \Bigg] \Bigg[ p^2\Bigg[ p^2\left( 1-\frac{1}{2}U\cdot V \right)
\nonumber \\&&
+(p\cdot U)(p\cdot V) \Bigg]\Bigg[ p^2\left( 1-\frac{1}{4}U^2V^2 \right)
\nonumber \\ &&
+\frac{1}{4}\Big( 4(p\cdot U)(p\cdot V)+(p\cdot V)^2U^2+(p\cdot U)V^2 \Big) \Bigg] \Bigg],
\end{eqnarray}
with the explicit expressions of the $F_i$ functions given by
\begin{eqnarray}
F_1( p )&=&-\frac{1}{2}\left( p^2 \right)^2\left( 1-\frac{1}{4}U^2V^2 \right)U\cdot V-\frac{1}{4}p^2(p\cdot U)(p\cdot V)\left( U^2V^2+2\,U\cdot V \right)
\nonumber \\ &&
+\frac{1}{4}\Big[ V^2(p\cdot U)^2+U^2(p\cdot V)^2 \Big]\left[ p^2\left( 1-\frac{1}{2}U\cdot V \right)+(p\cdot U)(p\cdot V) \right]
\nonumber \\  &&
+(\xi-1)\left[ -\frac{1}{2}\,p^2\,U\cdot V+(p\cdot U)(p\cdot V) \right]\bigg[ p^2\left( 1-\frac{1}{4}U^2V^2 \right)
\nonumber \\&&
+\frac{1}{4}\left( 4(p\cdot U)(p\cdot V)+(p\cdot V)^2U^2+(p\cdot U)^2V^2 \right) \bigg],
\label{F1}
\\ \nonumber \\
F_2( p )&=&-\frac{1}{2} p^2 \big[ p^2+ (p \cdot U)(p \cdot V)/2\big],
\\
F_3( p )&=&\frac{1}{2}\big[ (p \cdot V) p^2+(p \cdot U)(p \cdot V)^2-(p \cdot U)p^2V^2/2\big],
\\
F_4( p )&=&\frac{1}{2}\big[ (p \cdot U) p^2+(p \cdot V)(p \cdot U)^2-(p \cdot V)p^2U^2/2\big],
\\
F_5( p )&=&\frac{1}{4}p^2\big[ p^2 V^2-(p \cdot V)^2\big],
\\
F_6( p )&=&\frac{1}{4}p^2\big[ p^2 U^2-(p \cdot U)^2\big]. 
\end{eqnarray}
The definition 
\begin{equation}
\label{thetauv}
\Theta_{U,V}=-\frac{1}{2}U\cdot V+\frac{(p\cdot U)(p\cdot V)}{p^2}
\end{equation}
has also been utilized. Notice that invariance of Eq.~(\ref{2tensor}) with respect to the interchange of $U^\mu$ and $V^\mu$ has been inherited by the functions $\Theta_{U,V}$ and $\Xi^{\mu\nu}$. It is worth emphasizing that these functions trivially vanish in the limit in which $V_\mu\to0$ and $U_\mu\to0$, and consequently the low--energy photon propagator is consistently recovered. 

The modified propagator given in Eq.~(\ref{propagator}), which is the main ingredient for our calculation of the one--loop Lorentz--violating corrections to the $ff\gamma$ vertex, has an involved structure, so that, in order to simplify our derivation, we restrict our framework and obtain a simpler expression of this object. The first simplification that we consider is the election of a specific gauge, which we choose to be the Feynman--'t Hooft gauge ($\xi=1$). Notice that the $F_1$ function, Eq.~(\ref{F1}), has been written in such a way that the choice of this gauge does not affect its first three terms, while the cancelation of the last one is explicit. It is also worth mentioning that the photon propagator exhibited here coincides, in this gauge, with the expression previously reported in the literature~\cite{CFGdS}.
The effective field theory description provided by the SME parametrizes the Lorentz--violating effects of a presumable fundamental theory whose characteristic energy scale might be the Planck mass. As there is a difference of seventeen orders of magnitude between the Planck mass and the electroweak scale, the coefficients of the SME are expected to be tiny. The restrictive bounds~\cite{KR} on the different coefficients of the minimal SME also indicate that the effects of violation of Lorentz invariance are small. For these reasons, a perturbative treatment is well--founded. A further simplification of the photon propagator can then be attained by keeping only effects of Lorentz--symmetry violation at the first order. It is important noticing that quadratic products of the $V^\mu$ and $U^\mu$ four--vectors are~\cite{CFGdS} parameters at the first order in the $k^{\mu\nu}$ tensor, and hence represent the first--order contributions to the photon propagator. Fixing the gauge to $\xi=1$ and omitting all new--physics effects but the first--order Lorentz--violating terms, we write the photon propagator as
\begin{eqnarray}
-i\tilde{\Delta}^{\mu\nu}( p )&\approx&-\frac{g^{\mu\nu}}{p^2}+\frac{1}{2p^2}\left[ U^\mu V^\nu+U^\nu V^\mu \right]
-\frac{p\cdot V}{2\left( p^2 \right)^2}\left[ p^\mu U^\nu+p^\nu U^\mu \right]-\frac{p\cdot U}{2\left( p^2 \right)^2}\left[ p^\mu V^\nu+p^\nu V^\mu \right],
\label{apropagator}
\end{eqnarray}
which is symmetric under the interchange of $U^\mu$ and $V^\mu$. This is the expression that we will employ for our subsequent calculations.

\section{Lorentz--violating contributions to the $ff\gamma$ vertex at the one--loop level}
\label{radiative}
\begin{figure}[!ht]
\center
\includegraphics[width=7cm]{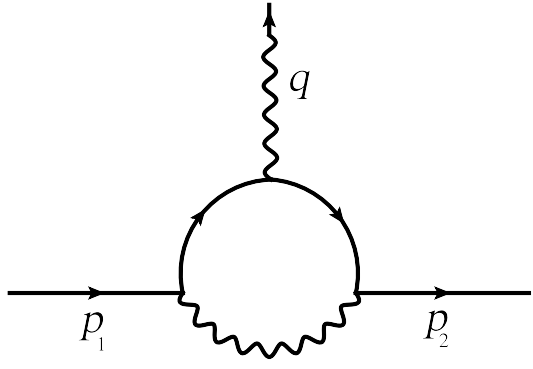}
\caption{\label{eegamma} The only diagram contributing to the one--loop $ee\gamma$ vertex with nonbirefringent Lorentz--violation.}
\end{figure}

In this section we derive one--loop contributions to the $ff\gamma$ vertex from the Lorentz--violating and $CPT$--even Maxwell sector of the minimal SME. We insert the modified photon propagator given in Eq.~(\ref{apropagator}) into the only contributing Feynman diagram, which is displayed in Fig.~(\ref{eegamma}). 
According to the conventions of Fig.~(\ref{eegamma}), the one--loop $ff\gamma$ vertex function, $\Gamma^\mu_f$, is given by
\begin{equation}
-ie\Gamma^\mu_f(q^2)=-ie^3\mu^{4-D}\int\frac{d^Dk}{(2\pi)^D}\,\tilde{\Delta}^{\alpha\beta}(k)\,\frac{\gamma_\alpha(\slashed{k}+\slashed{p}_2+m_f)\gamma^\mu(\slashed{k}+\slashed{p}_1+m_f)\gamma_\beta}{\Big[ (k+p_2)^2-m_f^2 \Big]\Big[ (k+p_1)^2-m_f^2 \Big]}.
\end{equation}
As it can be appreciated in this equation, we use the dimensional regularization scheme. In this context, the factor $\mu^{4-D}$, where $[\mu]={\rm mass}$, is introduced to correct units of the loop integral. We perform this calculation by using the Feynman--parameters technique, and we take the external fermions on shell, but assume that the external photon is off shell. The resulting expression can be written as the sum $\Gamma^\mu_f=\Gamma_{\rm SM}^\mu+\Gamma^\mu_{\rm SME}$, where $\Gamma_{\rm SM}^\mu$ is the SM contribution and $\Gamma^\mu_{\rm SME}$ is the new--physics contribution that originated in the minimal SME. The term $\Gamma^\mu_{\rm SME}$ has a complex structure, for several new couplings add to the well--known Lorentz--invariant structure~\cite{HIRSS}. It is given by
\begin{eqnarray}
\Gamma_{\rm SME}^\mu&=& G_1\,(U\cdot V)\,\gamma^\mu
+(G_{\rm uv}+G_2)(U_\mu\slashed{V}+V_\mu\slashed{U})
+\frac{1}{m_f}\,G_3\,(U\cdot V)\,i\sigma_{\mu\nu}q^\nu
\nonumber \\&&
+\frac{1}{m_f}G_4\Big[ V_\mu\,U\cdot(p_1+p_2)+U_\mu\,V\cdot(p_1+p_2) \Big]
\nonumber \\&&
+\frac{1}{m_f}G_5\Big[ V_\mu\,(U\cdot q)+U_\mu\,(V\cdot q)-(U\cdot q)\slashed{V}\gamma_\mu-(V\cdot q)\slashed{U}\gamma_\mu \Big]
\nonumber \\&&
+\frac{1}{m_f^2}G_6\,(U\cdot V)\,q^2\gamma_\mu
+\frac{1}{m_f^2}\,G_7\,(U\cdot q)(V\cdot q)\,\gamma_\mu
\nonumber \\&&
+\frac{1}{m_f^2}G_{8}\Big[ (U\cdot p_1)(V\cdot p_1)+(U\cdot p_2)(V\cdot p_2) \Big]\gamma_\mu
\nonumber \\&&
+\frac{1}{m_f^2}G_{9}\Big[ (U\cdot q)\slashed{V}+(V\cdot q)\slashed{U} \Big]q_\mu
\nonumber \\&&
+\frac{1}{m_f^2}G_{10}\Big[ \Big( (p_1\cdot U)\,p_{1\mu}+(p_2\cdot U)\,p_{2\mu} \Big)\slashed{V}
\nonumber \\&&
+\Big( (p_1\cdot V)\,p_{1\mu}+(p_2\cdot V)\,p_{2\mu} \Big)\slashed{U} \Big],
\label{oneloopffg}
\end{eqnarray}
which is invariant under the interchange of $U^\mu$ and $V^\mu$. The explicit expressions of the dimensionless factors $G_{\rm uv}$ and $G_i$, which are given in terms of parametric integrals, can be found in \ref{app}. The factor $G_{\rm uv}$, which is located in the second term of Eq.~(\ref{oneloopffg}), is ultraviolet--divergent, whereas all factors $G_i$ are free of such sort of divergences. As we will see below, the first and second terms of Eq.~(\ref{oneloopffg}) are the only ones that are generated, from the point of view of effective field theory, by renormalizable interactions. For that reason they are the only ones in which ultraviolet divergences are allowed to appear, although only the second term involves such sort of divergences. Note that ultraviolet divergences are consistently absent in all other terms, which can be produced only by nonrenormalizable interactions.

Infrared divergences~\cite{BN} occur in the calculation of loop contributions to the $ff\gamma$ vertex. For instance, the SM quantum electrodynamics contributions to electron scattering through the one--loop $ff\gamma$ vertex give rise to this sort of divergences. This problem is solved at the level of cross section by assuming that the photon has a small mass. Then, the inclusion of soft--bremsstrahlung diagrams generates logarithmic terms that cancel infrared divergences in the cross section. In the context of the SME, infrared divergences are generated when the fermionic propagator is modified by Lorentz--violating effects. The one--loop contributions to the $ff\gamma$ vertex that are produced by the renormalizable $CPT$--odd axial coupling of a vectorial background field to SM fermions were calculated~\cite{MNTT,smerc3} up to the second order in violation of Lorentz invariance, which yielded a contribution to the anomalous magnetic moment of fermions. It was found that most of these contributions, including those associated to the anomalous magnetic moment, involve infrared divergences that do not cancel, even in the cross section~\cite{smerc3}. This was then used to argue that this Lorentz--violating axial coupling is unphysical. This situation contrasts with the results found in the present paper, since most factors $G_i$, including the one that is related to the anomalous magnetic moment, are free of infrared divergences. The factors $G_i$ of the first, sixth, seventh, and eighth terms of Eq.~(\ref{oneloopffg}) are the only ones that contain infrared divergences and are all associated to the Dirac structure $\gamma_\mu$. This is what occurs in the case of the SM in the sense that the one--loop contributions to the vector coupling of the $ff\gamma$ vertex are the only ones that involve this sort of divergences. As we will show below, a finite contribution to the anomalous magnetic moment of fermions can be derived from the third term of Eq.~(\ref{oneloopffg}).

\begin{figure}[!ht]
\center
\includegraphics[width=7cm]{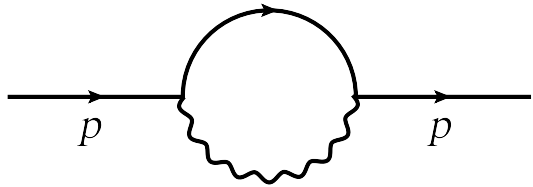}
\caption{\label{tpvf} One--loop diagram contributing to the $ff$ two--point function.}
\end{figure}

The presence of an external gauge boson in the $ff\gamma$ vertex sets the requirement of gauge invariance of these loop contributions.
It happens that imposing the on--shell conditions $q^2=0$ and $q_\mu\to0$ on Eq.~(\ref{oneloopffg}), the contraction of the one--loop vertex function $\Gamma^\mu_{\rm SME}$ with the momentum of the external photon is nonzero, that is
\begin{eqnarray}
q_\mu\Gamma^\mu_{\rm SME}&=&\frac{\alpha}{8\pi}\left( \Delta_\epsilon-{\rm log}\left( \frac{m_f^2}{\mu^2} \right)+2 \right)\Big( (p_1\cdot U)\,\slashed{V}+(p_1\cdot V)\,\slashed{U}
-(p_2\cdot U)\,\slashed{V}-(p_2\cdot V)\,\slashed{U} \Big)
\nonumber \\&&
+\frac{\alpha}{4\pi}\frac{1}{m_f}\Big( (p_1\cdot U)(p_1\cdot V)-(p_2\cdot U)(p_2\cdot V) \Big).
\label{wt1}
\end{eqnarray}
Nevertheless, a Ward--Takahashi identity is fulfilled instead. The diagram shown in Fig.~\ref{tpvf}
is the only one that contributes to the $ff$ two--point vertex function at the one--loop level. The calculation of this diagram, along with the usage of the modified photon propagator given in Eq.~(\ref{apropagator}), produces a one--loop amplitude, $\Sigma_f(p)$, that can be expressed as the sum of the SM contribution, $\Sigma_{\rm SM}(p)$, and a term $\Sigma_{\rm SME}(p)$, generated by the quantum electrodynamics sector of the minimal SME. The $\Sigma_{\rm SME}(p)$ function is given by
\begin{eqnarray}
-i\,\Sigma_{\rm SME}(p)&=&-i\Bigg[\frac{\alpha}{8\pi}\left( \Delta_\epsilon-{\rm log}\left( \frac{m_f^2}{\mu^2} \right)+2 \right)\Big( (p\cdot U)\,\slashed{V}+(p\cdot V)\,\slashed{U} \Big)
+\frac{\alpha}{4\pi}\frac{1}{m_f}\,(p\cdot U)(p\cdot V)
\nonumber \\&&
+\frac{\alpha}{8\pi}\,m_f\,\left( \Delta_\epsilon-{\rm log}\left( \frac{m_f^2}{\mu^2} \right)+1 \right)(U\cdot V)
\Bigg],
\end{eqnarray}
where the external fermions have been taken on shell. Using this expression, Eq.~(\ref{wt1}) can be rewritten as
\begin{equation}
q_\mu \Gamma^\mu_{\rm SME}=\Sigma_{\rm SME}(p_1)-\Sigma_{\rm SME}(p_2).
\end{equation}

Despite the contraction $q_\mu\Gamma^\mu_{\rm SME}$ is not equal to zero, this Ward--Takahashi identity indicates that the Lorentz--violating contribution $\Gamma^\mu_{\rm SME}$ is gauge invariant. Furthermore, consider the effective Lagrangian
\begin{eqnarray}
{\cal L}_{\rm eff}&=&\alpha^{(4)}_1\,(U\cdot V)\,i\bar{\psi}\gamma^\mu D_\mu\psi+\alpha^{(4)}_2(U^\mu V^\nu+U^\nu V^\mu)\,i\bar{\psi}\gamma_\mu D_\nu\psi
\nonumber \\&&
+\frac{\alpha^{(5)}_1}{m_f}(U\cdot V)\,\bar{\psi}\sigma^{\mu\nu}\psi F_{\mu\nu}
+\frac{\alpha^{(5)}_2}{m_f}\,(U^\mu V^\nu+U^\nu V^\mu)\,\bar{\psi}D_\mu D_\nu\psi
\nonumber \\ &&
+\frac{\alpha^{(5)}_3}{m_f}(U^\mu V^\nu+U^\nu V^\mu)\,\bar{\psi}\sigma_\nu\hspace{0.001cm}^\rho\psi F_{\mu\rho}
+\frac{\alpha^{(6)}_1}{m_f^2}(U\cdot V)\,\bar{\psi}\gamma^\nu\psi\,\partial^\mu F_{\mu\nu}
\nonumber \\&&
+\frac{\alpha^{(6)}_2}{m_f^2}(U^\mu V^\nu+U^\nu V^\mu)\bar{\psi}\gamma^\rho\psi\,\partial_\mu F_{\nu\rho}
+\frac{\alpha^{(6)}_3}{m_f^2}(U^\mu V^\nu+U^\nu V^\mu)\,i\bar{\psi}\gamma^\rho D_\mu D_\rho D_\nu\psi
\nonumber \\&&
+\frac{\alpha^{(6)}_4}{m_f^2}(U^\mu V^\nu+U^\nu V^\mu)\,\bar{\psi}\gamma_\mu\psi\,\partial^\rho F_{\rho\nu}
+\frac{\alpha^{(6)}_5}{m_f^2}(U^\mu V^\nu+U^\nu V^\mu)\,i\bar{\psi}\gamma_\mu D^\rho D_\nu D_\rho\psi,
\label{efflagr}
\end{eqnarray}
which is governed by the electromagnetic gauge group. This Lagrangian is built of renormalizable and nonrenormalizable terms with mass dimensions ranging from four to six, whose structures involve the Lorentz--violating four--vectors $U^\mu$ and $V^\mu$. Each of these terms also incorporate an unknown dimensionless coefficient $\alpha^{(i)}_j$. Any nonrenormalizable operator of ${\cal L}_{\rm eff}$ is divided by a power of the fermionic mass that corrects its units, so that the mass dimension of the whole term is four. Note that all these terms produce tree--level contributions to the $ff\gamma$ vertex.

Assuming that the fermions in the $ff\gamma$ interaction are on shell and the photon is off shell, the effective Lagrangian ${\cal L}_{\rm eff}$ generates the vertex function
\begin{eqnarray}
\Gamma_{\rm eff}^\mu&=&\alpha^{(4)}_1(U\cdot V)\,\gamma_\mu+\alpha^{(4)}_2(U_\mu\slashed{V}+V_\mu\slashed{U})-\frac{\alpha^{(5)}_1}{m_f}\frac{2}{e}\,(U\cdot V)\,i\sigma_{\mu\nu}q^\nu
\nonumber \\&&
+\frac{\alpha^{(5)}_2}{m_f}\Big[ V_\mu\,U\cdot(p_1+p_2)+U_\mu\,V\cdot(p_1+p_2) \Big]
\nonumber \\&&
+\frac{\alpha^{(5)}_3}{m_f}\frac{1}{e}\bigg[ V_\mu(U\cdot q)+U_\mu(V\cdot q)-\Big((U\cdot q)\slashed{V}\gamma_\mu+(V\cdot q)\slashed{U}\gamma_\mu\Big)
\nonumber \\&&
+2m_f(U_\mu\,\slashed{V}+V_\mu\slashed{U})
-\Big(V_\mu\,U\cdot (p_1+p_2)+U_\mu V\cdot(p_1+p_2)\Big) \bigg]
\nonumber \\&&
+\frac{\alpha^{(6)}_1}{m_f^2}\frac{1}{e}\,(U\cdot V)\,q^2\gamma_\mu+\frac{\alpha^{(6)}_2}{m_f^2}\frac{2}{e}(U\cdot q)(V\cdot q)\gamma_\mu
\nonumber \\&&
+\frac{\alpha^{(6)}_3}{m_f^2}\bigg[ -m_f\Big(V_\mu\,U\cdot(p_1+p_2)+U_\mu\,V\cdot(p_1+p_2) \Big)+(U\cdot q)(V\cdot q)\gamma_\mu
\nonumber \\&&
-\Big( (U\cdot p_1)(V\cdot p_1)+(U\cdot p_2)(V\cdot p_2) \Big)\gamma_\mu
\bigg]
\nonumber \\&&
+\frac{\alpha^{(6)}_4}{m_f^2}\frac{1}{e}\Big[ (U_\mu\slashed{V}+V_\mu\slashed{U})q^2-\Big( (U\cdot q)\slashed{V}+(V\cdot q)\slashed{U} \Big)q_\mu \Big]
\nonumber \\&&
+\frac{\alpha^{(6)}_5}{m_f^2}\bigg[ \frac{q^2-2m_f^2}{2}(U_\mu\slashed{V}+V_\mu\slashed{U})-\Big( (U\cdot p_1)\,p_{1\mu}+(U\cdot p_2)\,p_{2\mu} \Big)\slashed{V}
\nonumber \\&&
-\Big( (V\cdot p_1)\,p_{1\mu}+(V\cdot p_2)\,p_{2\mu} \Big)\slashed{U} \bigg].
\end{eqnarray}
Since ${\cal L}_{\rm eff}$ is invariant with respect to the electromagnetic gauge group, the vertex function $\Gamma_{\rm eff}^\mu$, which contains all tree--level contributions from this effective Lagrangian to the $ff\gamma$ interaction, also possesses this symmetry. It can be straightforwardly verified that the one--loop vertex function $\Gamma^\mu_{\rm SME}$, Eq.~(\ref{oneloopffg}), can be written in this form, which is another way to prove that $\Gamma^\mu_{\rm SME}$ is gauge invariant.

The profit of employing the Lorentz--violating effective Lagrangian ${\cal L}_{\rm eff}$ is twofold: besides being useful for the analysis of gauge invariance of the Lorentz violating vertex function $\Gamma_{\rm SME}^\mu$, the effective Lagrangian ${\cal L}_{\rm eff}$ and, consequently, its tree--level trilinear vertex function $\Gamma_{\rm eff}^\mu$ carry information about the transformation properties of the terms in such loop contribution under $C$, $P$, $T$ and its combinations. The transformation properties under $C$, $P$, $T$, $CP$, and $CPT$ of the effective terms of the Lagrangian ${\cal L}_{\rm eff}$, Eq.~(\ref{efflagr}), are shown in Table~\ref{tab1}.
\begin{table}[ht]
\centering
\begin{tabular}{| c | c | c | c | c | c |}
\hline
& $C$ & $P$ & $T$ & $CP$ & $CPT$ 
\\ \hline
$\alpha_i^{(j)} (U\cdot V)$ & + & + & + & + & +
\\[0.2cm]
$\alpha_i^{(j)}(U^0V^0+U^0V^0)$ &+ & + & + & + & +
\\[0.2cm]
$\alpha_i^{(j)}(U^0V^j+U^jV^0)$ & + & \textminus & \textminus & \textminus & +
\\[0.2cm]
$\alpha_i^{(j)}(U^jV^k+U^kV^j)$ &+ & + & + & + & +
\\ \hline
\end{tabular}
\caption{\label{tab1}Transformation properties of nonrenormalizable operators in the effective Lagrangian ${\cal L}_{\rm eff}$ under $C$, $P$, $T$, $CP$, and $CPT$.}
\end{table}
The $\alpha^{(i)}_j (U\cdot V)$, placed in the first line of this table, represents any effective term of ${\cal L}_{\rm eff}$ that contains the Lorentz scalar $(U\cdot V)$. All other terms in the Lagrangian have the form
\begin{equation}
\alpha^{(j)}_i(U^\mu V^\nu +U^\nu V^\mu)\,T_{\mu\nu}=\alpha^{(j)}_i(U^0V^0+U^0V^0)\,T_{00}
+\alpha^{(j)}_i(U^0V^j+U^jV^0)(T_{0j}+T_{j0})
+\alpha^{(j)}_i(U^jV^k+U^kV^j)T_{jk},
\end{equation}
in which the first term has only timelike components of the four--vectors $U^\mu$ and $V^\nu$, the third term is exclusively proportional to spacelike components of such four--vectors, and the second term involves a sum of products of a timelike component with a spacelike  component. Terms associated to $U^0V^0+U^0V^0$ or $U^j V^k+U^kV^j$ are even under any discrete transformation $C$, $P$, or $T$, whereas terms involving factors $U^0V^j+U^jV^0$ are even under $C$, but odd with respect to $P$, $T$, and $CP$. In all cases $CPT$ is preserved, as it occurs~\cite{smerc5} in the case of the renormalizable Lorentz--violating interaction in the Maxwell sector.

According to the effective Lagrangian ${\cal L}_{\rm eff}$ and the effective vertex function $\Gamma^\mu_{\rm eff}$, the Lorentz--violating one--loop vertex function $\Gamma^\mu_{\rm SME}$ contains contributions to the vector current (first term), to the anomalous magnetic moment (third term), and to the anapole moment (sixth term), which are low energy observables already present in the usual Lorentz--preserving parametrization~\cite{HIRSS} of the $ff\gamma$ vertex. All these contributions are proportional to the Lorentz scalar $(U\cdot V)$, so they are consistently Lorentz invariant.
On the other hand, terms involving this Lorentz scalar are even under $C$, $P$, and $T$. This suggests that, in general, in a given calculation involving this sort of Lorentz--violating effects, terms
that are proportional to the Lorentz scalar $(U\cdot V)$ are the only ones whose space--time symmetry properties would not forbid contributions to low--energy observables. It is important keeping in mind that this discussion only makes sense in the case of one--loop contributions at the first order in violation of Lorentz invariance.

As we commented before, the one--loop vertex function $\Gamma^\mu_{\rm SME}$ contains a new--physics contribution to the anomalous magnetic moment of fermions, which we denote by $a^{\rm SME}_f$. The anomalous magnetic moments of the electron and the muon have been calculated, in the SM, with remarkable accuracy~\cite{tsmamm1,tsmamm2,tsmamm3}. At present the disagreements between the experimentally measured values~\cite{eamm1,eamm2,eamm3,eamm4} and the theoretical predictions of these physical quantities are~\cite{tsmamm1,tsmamm2,tsmamm3} $\Delta a_e=a_e^{\rm EXP}-a_e^{\rm SM}=-1.06\times 10^{-12}$ and $\Delta a_\mu=a^{\rm EXP}_\mu-a^{\rm SM}_\mu=2.49\times10^{-9}$, which makes it a place where suppressed new physics could manifest. Employing the usual parametrization of the Lorentz--invariant $ff\gamma$ vertex~\cite{HIRSS}, we write the contribution from the $\Gamma^\mu_{\rm SME}$ one--loop vertex function as
\begin{equation}
a_f^{\rm SME}=-\frac{\alpha}{4\pi}\,U\cdot V,
\end{equation}
which we obtained by taking $q^2=0$ in the magnetic dipole form factor emerged from the third term of Eq.~(\ref{oneloopffg}). Now we concentrate in the contributions from isotropic violation of Lorentz invariance, so we take $U^\mu=(U^0,0,0,0)$ and $V^\mu=(V^0,0,0,0)$, which are invariant under spatial rotations. Notice that the $U^0$ and $V^0$ components are subjected to fulfill
\begin{equation}
\label{condition}
U^0V^0=2\,\tilde{\kappa}_{\rm tr}.
\end{equation}
For instance, in Ref.~\cite{twlv10} a modified photon propagator that includes only isotropic effects of Lorentz violation was derived. The authors of that work pointed out that their result matches the one of Casana {\it et al}.~\cite{CFGdS} for $V^\mu=U^\mu=(\sqrt{2\tilde{\kappa}_{\rm tr}},0,0,0)$, which satisfy this condition. Taking Eq.~(\ref{condition}) into account, we note that $U\cdot V=2\,\tilde{\kappa}_{\rm tr}$, which we utilize to write the magnetic dipole contribution $a^{\rm SME}_f$ as
\begin{equation}
a^{\rm SME}_f=-\frac{\alpha}{2\pi}\,\tilde{\kappa}_{\rm tr}.
\end{equation}
This means that the total contribution to anomalous magnetic moment, $a_f$, from both the SM, $a^{\rm SM}_f$, and the minimal SME, $a^{\rm SME}_f$, has the form $a_f=(1-\tilde{\kappa}_{\rm tr})\,a^{\rm SM}_f$. This result exactly coincides with the one derived in Ref.~\cite{smerc7} by insertion of a two--point function carrying Lorentz--violating effects from the $CPT$--even pure--photon sector of the SME into the one--loop diagram. Since the difference between the SM theoretical prediction and the experimentally measured value of the anomalous magnetic moment of the electron has been reduced by one order of magnitude with respect to the data that was available when Ref.~\cite{smerc7} was published, the corresponding bound on $\tilde{\kappa}_{\rm tr}$ can be updated to
\begin{equation}
\tilde{\kappa}_{\rm tr}\lesssim9\times10^{-10},
\end{equation}
which is weaker than the most stringent bounds currently available~\cite{KR}. The constraint set by the anomalous magnetic moment of the muon is even less restrictive.

\section{Conclusions}
\label{conc}
In the present paper, we calculated the one--loop contributions to the $ff\gamma$ vertex that are generated by a photon propagator carrying modifications induced by nonbirefringent $CPT$--even and Lorentz--violating effects from the coupling $(K_F)^{\mu\nu\alpha\beta}$, which is part of the minimal SME. The corresponding expression is given in terms of two four--vectors, $U^\mu$ and $V^\mu$. We considered such modified propagator and restricted it to the first order in Lorentz violation. We then utilized the resulting expression to calculate the only one--loop diagram contributing to the $ff\gamma$ vertex, with the external fermions on shell and the external photon off shell, and found a rich gauge structure. In order to prove gauge invariance of our result, we contracted the momentum of the external photon with the resulting one--loop $ff\gamma$ amplitude evaluated on shell. We found that the result of this operation does not yield a simple Ward identity. Instead, a Ward--Takahashi identity is obtained, which proves that our result is gauge invariant. We provided another proof which consisted in constructing a set of gauge invariant nonrenormalizable operators generating, at tree level, the same structures that we obtained in our derivation of the one--loop $ff\gamma$ vertex. Since the structure of the nonrenormalizable operators is governed by the electromagnetic gauge symmetry group, their tree--level contributions to the $ff\gamma$ vertex function are gauge invariant, and, therefore, our loop expression is also gauge invariant. The properties of the nonrenormalizable terms with respect to the discrete transformations $C$, $P$, and $T$, gave us information about the behavior under such transformations of the one--loop $ff\gamma$ vertex that we calculated. Particularly, we found that, at the first order in Lorentz violation, terms involving the Lorentz scalar $(U\cdot V)$ are the only ones whose properties under space--time transformations and charge conjugation meet the necessary requirements to contribute to low--energy observables. Finally, we calculated the contribution to anomalous magnetic moment of fermions from this loop vertex and found agreement with the expression previously reported in the literature, which was calculated through a perturbative insertion carrying Lorentz violation.

\acknowledgments{We acknowledge financial support from CONACYT and SNI (M\' exico).}

\appendix

\section{List of parametric integrals}
\label{app}
\begin{eqnarray}
G_1&=&\frac{\alpha}{4\pi}\int_0^1dx\int_0^{1-x}dy\,\Bigg[(3x+3y-2)\,{\rm log}\left( \frac{m_f^2 (x+y)^2-q^2 x y}{\mu^2} \right)
\nonumber \\&&
+(x+y)\Bigg( 1
+\frac{m_f^2 \left(x^2+2 x (y-3)+(y-2)
   y+2\right)}{m_f^2
   (x+y)^2-q^2 x y} \Bigg)\Bigg],
\\ \nonumber \\
G_{\rm uv}&=&\frac{\alpha}{4\pi}\int_0^1dx\int_0^{1-x}dy\,\left[ \Delta_\epsilon-{\rm log}\left( \frac{m_f^2 (x+y)^2-q^2 x y}{\mu^2} \right) \right],
\\ \nonumber \\
G_2&=&-\frac{\alpha}{4\pi},
\\ \nonumber \\ 
G_3&=&-\frac{\alpha m_f^2}{2\pi}\int_0^1dx\int_0^{1-x}dy\frac{x (x+y)}{m_f^2 (x+y)^2-q^2 x y}
\\ \nonumber \\
G_4&=&\frac{\alpha m_f^2}{8\pi}\int_0^1dx\int_0^{1-x}dy\,(x+y)^2 \Bigg[\frac{-4 x-4 y+6}{m_f^2 (x+y)^2-q^2 x
   y}
\nonumber \\&&
   +\frac{(x+y-1) \left(m_f^2 (x+y-2) (x+y)+q^2
   (x-x y)\right)}{\left(m_f^2 (x+y)^2-q^2 x
   y\right){}^2}\Bigg]
\\ \nonumber \\
G_5&=&\frac{\alpha m_f^2}{8\pi}\int_0^1dx\int_0^{1-x}dy\,(x+y)\Bigg[ \frac{q^2 (x-y)^2 (x+y-1) }{2 \left(m_f^2
   (x+y)^2-q^2 x y\right){}^2}
\nonumber \\ &&   
   -\frac{
   (x+y+1)}{m_f^2 (x+y)^2-q^2 x y} \Bigg]
\\ \nonumber \\ 
G_6&=&\frac{\alpha m_f^2}{8\pi}\int_0^1dx\int_0^{1-x}dy\,\frac{(x+y)(-2 x y+x+y-1)}{ m_f^2
   (x+y)^2- q^2 x y}
\\ \nonumber \\
G_7&=&\frac{\alpha m_f^2}{4\pi}\int_0^1dx\int_0^{1-x}dy\,(x+y-1)\Bigg[ -\frac{4 (x (2 y-1)-y)}{m_f^2 (x+y)^2-q^2 xy}
\nonumber \\&&   
+\frac{\left(m_f^2 (x+y-2) (x+y) (x
   (2 y-1)-y)-2 q^2 (x-1) x (y-1)
   y\right)}{\left(m_f^2 (x+y)^2-q^2 x
   y\right){}^2} \Bigg]
   \end{eqnarray}
\begin{eqnarray}
G_{8}&=&\frac{\alpha m_f^2}{4\pi}\int_0^1dx\int_0^{1-x}dy\,(x+y-1) \Bigg[\frac{2 x (4 x+4 y-5)-4 y}{m_f^2
   (x+y)^2-q^2 x y}
\nonumber \\&&
-\frac{(x+y)}{\left(m_f^2 (x+y)^2-q^2 xy\right){}^2} \Big(m_f^2 (x+y-2)
   (x (2 x+2 y-3)-y)
\nonumber \\&&
   -2 q^2 (x-1) x
   (y-1)\Big)\Bigg]
\end{eqnarray}

\begin{eqnarray} 
G_{9}&=&-\frac{\alpha m_f^2}{4\pi}\int_0^1dx\int_0^{1-x}dy\,y\Bigg[ \frac{ x (4 x+4 y-3)-y+1}{m_f^2 (x+y)^2-q^2 x
   y}
\nonumber \\&&   
+\frac{x  (-x-y+1) \left(m_f^2 (x+y-2)
   (x+y)+q^2 (y-x y)\right)}{\left(m_f^2
   (x+y)^2-q^2 x y\right){}^2} \Bigg]
\\ \nonumber \\
G_{10}&=&\frac{\alpha m_f^2}{4\pi}\int_0^1dx\int_0^{1-x}dy\,(x+y-1) \Bigg[\frac{4 x (x+y)-y}{m_f^2 (x+y)^2-q^2
   x y}
\nonumber \\&&
   +\frac{x (x+y) \left(q^2 (x-1) y-m_f^2
   (x+y-2) (x+y)\right)}{\left(m_f^2 (x+y)^2-q^2 x
   y\right){}^2}\Bigg]
\end{eqnarray}

\end{document}